# Mass Production of a Trigger Data Serializer ASIC for the Upgrade of the Muon Spectrometer at the ATLAS Experiment

Jinhong Wang, Xiong Xiao, Reid Pinkham, Liang Guan, Wenhao Xu, Zhongyao Qian, Prachi Arvind Atmasiddha, Jacob Searcy, John Chapman, Bing Zhou , and Junjie Zhu

*Department of Physics, University of Michigan, Ann Arbor, MI, 48109, USA*

A B S T R A C T

The Trigger Data Serializer (TDS) is a custom ASIC designed for the upgrade of the innermost station of the endcap ATLAS Muon spectrometer. It is a mixed-signal chip with two working modes that can handle up to 128 detector channels. A total of 6,000 TDS ASICs have been produced for detector operation. This paper discusses a custom automatic test platform we developed to provide quality control of the TDS ASICs. We introduce the design, test procedures, and results obtained from this TDS testing platform.

## 1. Introduction

The present endcap innermost station of the ATLAS muon spectrometer (called the small wheel detector) will be replaced by a New Small Wheel (NSW) detector [1] to handle increased trigger and readout data rates expected at the high-luminosity Large Hadron Collider (HL-LHC). Small-strip Thin Gap Chambers (sTGC) [2] have been selected as one of the two detector technologies for the NSW upgrade along with Micromegas (MM) detectors. The two detector technologies are complementary. Both sTGC and MM detector will provide trigger and tracking primitives to the ATLAS trigger and readout system.

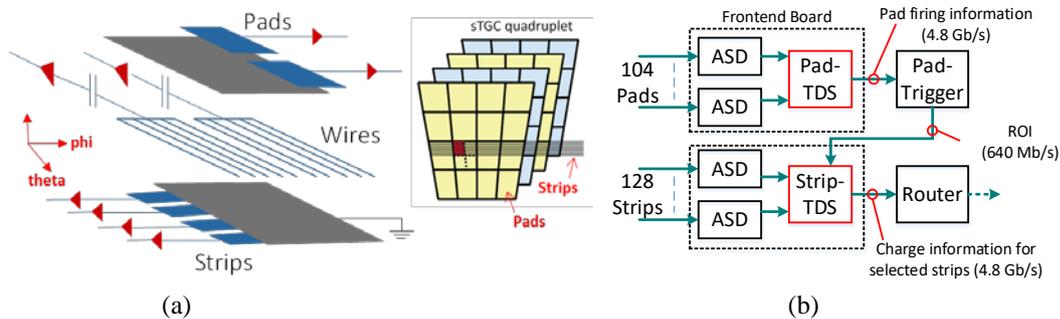

Fig. 1. (a) Basic structure of the sTGC detector and an illustration of four layers of pad-wire-strip planes in an sTGC quadruplet; (b) sTGC frontend electronics chain showing TDS ASICs. See text for explanation.

The basic structure of the sTGC detector is shown in Fig. 1(a). It is a multi-wire proportional drift chamber with a grid of gold-plated tungsten anode wires in a gas volume sandwiched between two resistive cathode planes. Charged particles ionize the gas creating electrons which drift to the wires and produce a current pulse on the wire. Signals induced from the wire are read on both sides of the anode plane: strips on one side provide a precision coordinate measurement and trapezoidal pads on the other side provide signals for fast triggering. Strips have a pitch of 3.2 mm and lengths of 1-2 m whereas pads vary in size from 30 $cm^2$ to 500 $cm^2$.

A diagram of the signal flow of the sTGC trigger chain is shown in Fig. 1(b). Raw detector signals from both pads and strips are first processed by a 64-channel Amplifier-Shaper-Discriminator (ASD) ASIC [3]. Digitized ASD outputs are sent to a Trigger Data Serializer (TDS) ASIC [4-5]. The TDS ASIC has two operating modes to handle the pad and strip detector information, denoted as pad-TDS and strip-TDS, respectively. The pad-TDS checks for the presence of pad detector signals, prepares the trigger data for up to 104 pads, and sends the data together with the LHC Bunching Crossing Identification number (BCID) to the pad-trigger board (Pad-Trigger) at a rate of 4.8 Gb/s. The pad-trigger board collects pad-TDS data from eight sTGC layers and determines a region-of-interest (ROI) for the candidate muon track. The ROI is then encoded and transmitted to the strip-TDS at a rate of 640 Mb/s. The strip-TDS decodes the deposited charge of all strips connected to the ASD and stores the charge information together with corresponding BCIDs in buffers. After receiving the ROI from the pad-trigger board, pad-strip matching is performed using a pre-assigned lookup table (LUT). Only charges from matched strips are packed and serialized to the signal router board (Router) on the rim of the NSW detector at a rate of 4.8 Gb/s [6-7].

The TDS ASIC was designed in a 130 nm CMOS technology. Mass production started in 2018, and in total about 6,000 TDS chips were produced. Thorough characterization of each ASIC is mandatory before they are mounted on frontend boards. A three-layer test platform was designed and fabricated to test all 6,000 chips. The platform was designed to provide a complete test environment for the TDS ASIC and characterize each TDS ASIC automatically with minimal user intervention. The test platform contains a hardware setup to emulate functions of companion circuits in the detector system; firmware implemented in a field-programmable gate arrays (FPGA) for stimulus and data processing; and software running on a PC for user control and communication. The platform is fully custom and scalable. Multiple setups can be configured and utilized in parallel to speed up the testing process.

This paper is arranged as follows. Section 2 gives an overview of the TDS ASIC, followed by an introduction of the test procedures for each part in the ASIC. Section 3 describes the test platform system and Section 4 has conclusions.

## 2. Verification Methodology for the TDS ASIC

A simplified block diagram of the TDS is shown in Fig. 2. The chip is divided into three parts according to the signal flow inside the chip: ASD interface, Preprocessor, and Serialization. Both pad and strip modes utilize a common configuration interface (I2C) and share the same serializer (GBT SER DM) [8] for the 4.8 Gb/s output.

In the pad mode, detector signals from fired pads are converted into pulses by the ASD and these pulses are captured by the "Pulse-detector" unit in each channel. A BCID time tag is also assigned in the presence of a pad pulse and is buffered in the corresponding channel ring buffer. The ring buffer is checked every 25 ns for the firing status of the corresponding pad. If the current BCID is found a "YES" of the firing status is asserted, otherwise a "NO" flag is attached. Each pad occupies a relatively large area, thus the routing from the detector to the ASD could introduce systematic variations in timing. There is a timing compensation circuit implemented in each channel to compensate these variations [5]. The delay is compensated by adjusting the phase of the timing clock (*CLK40*) instead of adding delay in the signal path. This is achieved by a phase programmable clock generator in each pad channel. An illustration of the compensation principle is shown in Fig. 3(a), in which a four-step phase programmable clock generator is illustrated with a 6.25 ns step. By making use of both leading and trailing edges of the 160 MHz clock (*CLK160*), a 3.125 ns phase step is obtained. Details on the circuit implementation can be found in [5]. For the quality control, the compensation scheme was evaluated using a statistical analysis. There are about 3580 BCs in a complete LHC orbit cycle. In each orbit cycle, we generate a pad pulse only at a particular BC (for example BCID *k*), and then subdivide BC *k* into 8 portions of size 3.125 ns, as shown in Fig. 3(b). A token circulates through all 8 subdivisions inside the BC *k* moving one subdivision per LHC orbit. A pulse is generated once every orbit from the time slot where the token is present. Verification begins by first adjusting the phases of the pad pulses so that they all fall into a single BCID (BCID *m*) when there is no compensated delay. Once this initial condition is established, the number of hits in the BCID *m* is expected to be reduced by 1/8 for every additional 3.125 ns compensation delay introduced, which is used as the criteria for performance assertion.

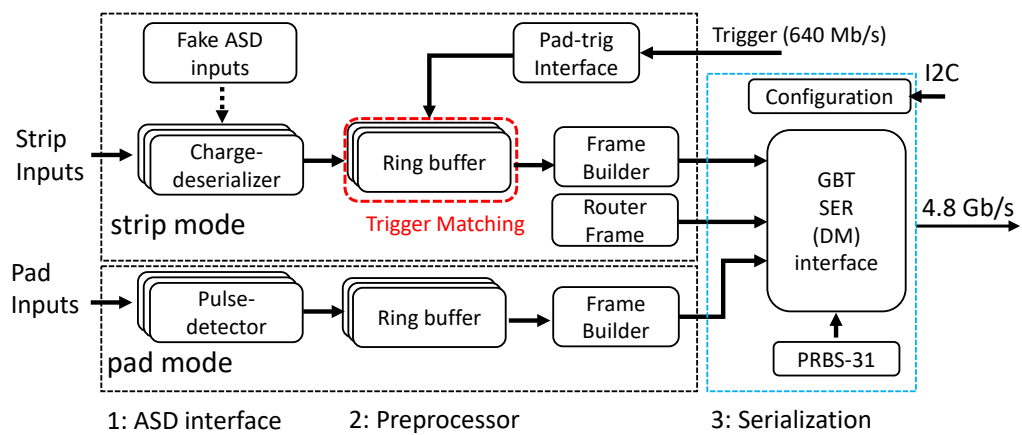

Fig. 2. Simplified signal flow of the pad and strip modes in TDS.

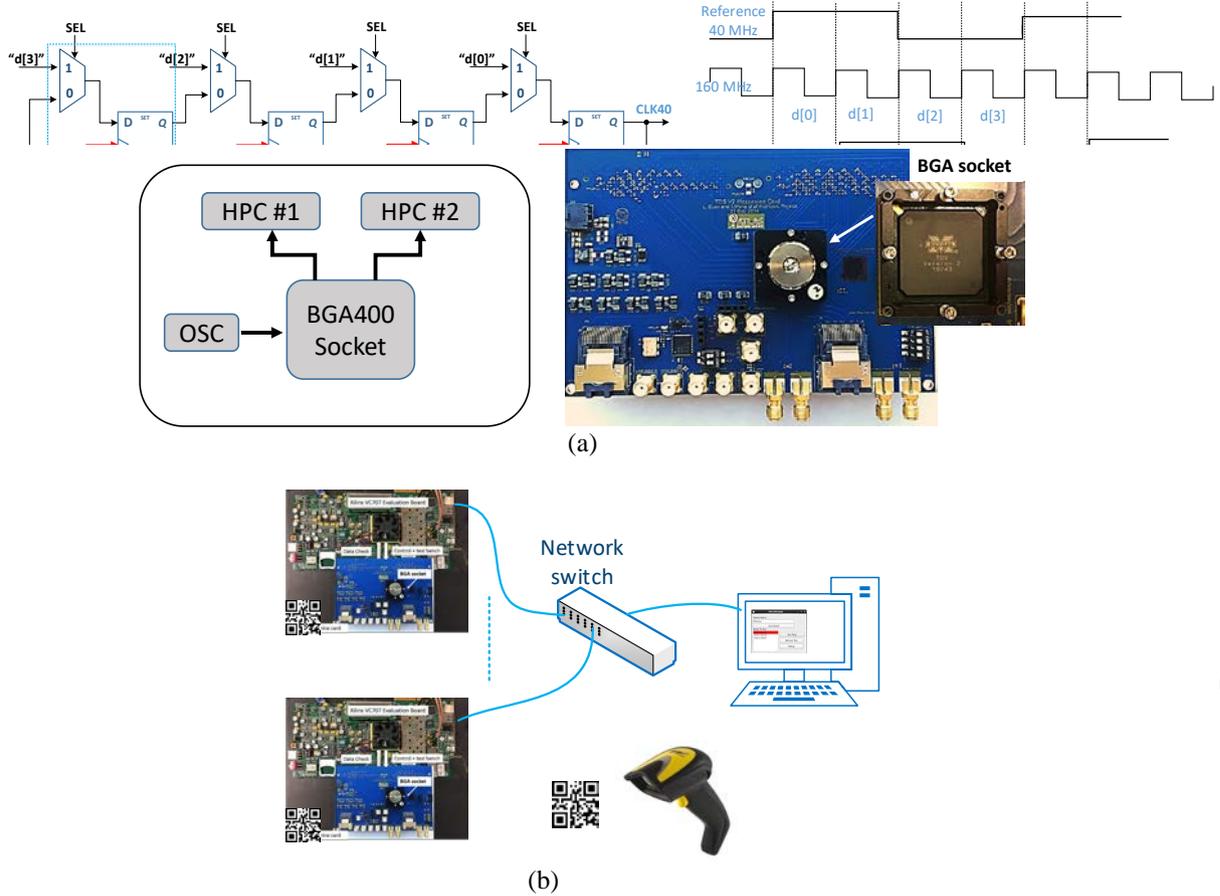

(a)

(b)

Fig. 4. (a) A block diagram of the mezzanine card and its picture (HPC connectors are on the other side); (b) Illustration of the test platform with multiple stations cascaded together. See text for explanation.

In the strip mode, strip charges sent by the ASD are captured by the "Charge-deserializer" unit and the decoded charges are stored in a ring buffer together with their BCID tags. When a trigger signal arrives, a band of strip channels are selected via a configurable LUT and the trigger matching is performed by checking the BCID tags. Strip charges having a BCID within a given time window together with the trigger information are collected and reformatted for serialization. Verification of strip-TDS is performed by configuring trigger LUTs to obtain a full coverage of the 128 channels. In addition, there are diagnosis functions in the strip mode: "bypass trigger" and "Router testing frame". In "bypass trigger", the external trigger signal can be bypassed and thus the signal path of each strip channel can be tested individually. There are also "fake ASD inputs" to the first 14 channels in case no external test input data is available. In "Router testing frame", data and NULL packets are emulated in a specific sequence to train the packet switching algorithm in the Router [7]. The connection between the strip-TDS and the Router is done via a 4.8 Gb/s serial link. The selected strip information from a trigger is transmitted in data packets. NULLs are inserted to keep the link running continuously when there are no data available.

The serialization interface is shared between the pad and strip modes, and a pseudo-random binary sequence generation with a permutation of 31 bits (PRBS-31 with a polynomial functional of $x^{31}+x^{28}+1$) is embedded for characterizing the performance of the serial link. Verification of the serialization interface is performed by checking the bit error ratio (BER) with the embedded PRBS-31. TDS is configured through an I2C interface, and its verification is done by writing all registers and reading them back for comparison.

## 3. Test Platform Setup and Test Procedure

### 3.1. Test Platform Hardware Setup

The TDS was packaged in a 400-pin Ball Grid Array (BGA) package. A mezzanine card mounted on a Xilinx VC707 evaluation board was designed for the chip performance evaluation. A 400-pin BGA socket from Ironwood Electronics [9] was utilized to hold the TDS ASIC. A block diagram of the mezzanine card is shown in Fig. 4(a), in which the reference clock is provided by an on-board oscillator (OSC). The mezzanine is attached to the VC707 board via a pair of High-Pin-Count (HPC) connectors, and they are referred together as a test station. Multiple tests in a test station are executed and analyzed in the VC707, and different test stations are coordinated by a software program running on a PC. The communication is done via Ethernet with User Datagram Protocol (UDP) running. Each test station is independent and multiple stations can be cascaded via a network switch to accelerate the production testing progress, as shown in Fig. 4(b).

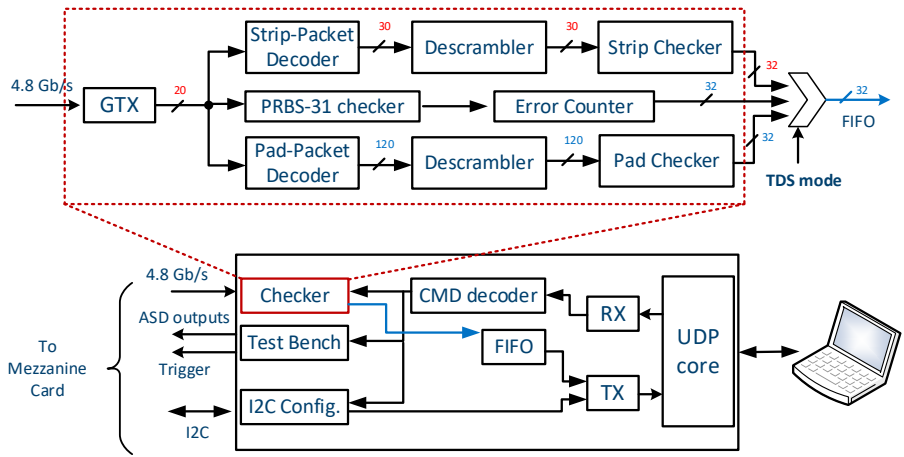

Fig. 5. The block diagram of the test firmware inside the FPGA on a VC707 evaluation board.

*3.2. Test Platform Firmware and Software*

The Virtex-7 FPGA is a core component on the VC707 evaluation board. It provides test inputs to the TDS, handles configuration controls, and processes the TDS 4.8 Gbps output data. Generation of input test data and checking of feedback data was done inside the FPGA to keep the data flow to the PC manageable. A block diagram of the firmware architecture is shown in Fig. 5. There are three major components: the Ethernet interface to the PC ("UDP core"), the I2C configuration block ("I2C Config."), and the TDS data checker ("Checker") together with the test bench generator ("Test Bench"). A TDS test starts from sending a configuration command through the software graphic user interface (GUI) to the Virtex-7 FPGA for a specific TDS configuration. The UDP core inside the FPGA decodes the command in the "CMD decoder" and passes relevant information to configure all three components. Once the I2C configuration is done, the TDS is set to a designated working mode, "Test Bench" is ready to generate emulated ASD pulses, and the "Checker" is waiting to analyze feedback packets. A start command is sent through the software GUI to start the test. Emulated ASD outputs are released and the checking status are streamed in the "FIFO" to be sent back to the PC. The test completes in response to a "stop" command from the software GUI.

The 4.8 Gb/s stream is reassembled in 20-bit groups by the GTX transceiver in the VC707 FPGA. The TDS packets in the strip mode are in 30-bit frames while the length of those in the pad mode is 120 bits. The 20-bit raw data is buffered to be rearranged in the right format and length. This is achieved by keeping track of the header of each frame: every 30-bit frame in the strip mode or the 120-bit frame in the pad mode starts with a 4-bit header, and the boundary of consecutive packets can be identified by checking the unique header patterns at the same position every 30 or 120 bits for the strip or pad mode, respectively. The payload following the headers is scrambled to keep the serial stream DC balanced. Once the boundary of a packet is identified, it is recovered by the "Descrambler" in the packet decoder. Depending on the TDS operation mode and specific test being run, descrambled packets are cross checked with the test input data accordingly. Corresponding checking summary is forwarded to the software GUI for performance evaluation. The serialization core is already included in the evaluation of the pad or strip mode, while a thorough characterization is available by feeding PRBS-31 patterns and checking the received stream for integrity. This is done by an embedded PRBS-31 generator inside the TDS and the "PRBS-31" checker inside the FPGA on the VC707 board as shown in Fig. 5.

*3.3. Test Procedure*

Verification of a TDS starts from preparing its test station and labeling the chips with QR codes. A TDS is picked up and placed into the BGA socket via a vacuum pen, as shown in Fig. 6(a). Electrostatic Discharge (ESD) can produce severe damage in the chips thus an anti-static wristband is required while handling chips. Once a TDS is inserted into a socket, QR codes of the TDS chip and the mezzanine card are scanned (Fig. 6(b)), and the chip is paired to its test station in the software GUI before closing the socket lid (Fig. 6(c)). The VC707 board is powered up and the FPGA is configured with preloaded firmware from the on-board flash memory which encodes the whole procedure shown in Fig. 5. In addition, the name of person conducting the test is recorded in the software GUI, so that the testing progress can be tracked. At the end of the chip test, results are summarized and a log file is created as shown in Fig. 6(d).

The software GUI takes control of the whole verification process by issuing test commands to each test station and collecting feedback status to navigate further evaluations. There are in total 6 tests, as shown in Fig. 7:

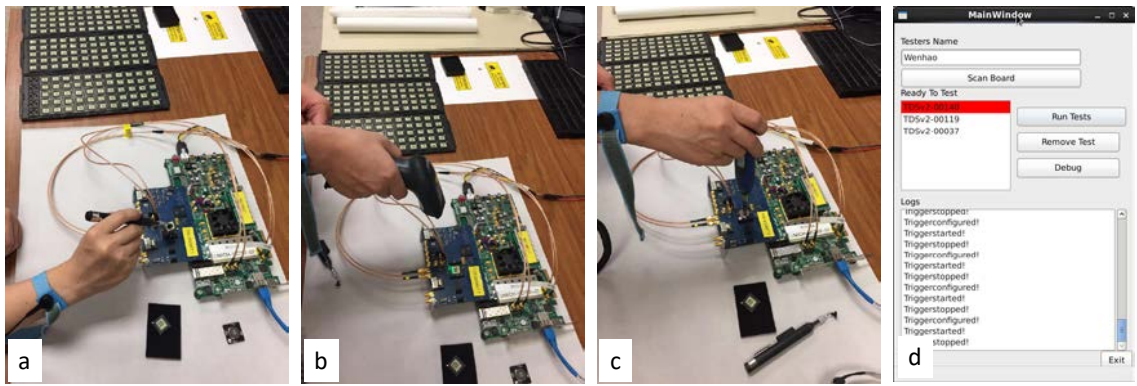

Fig. 6. Steps for preparing a station for the TDS test: (a) placement of the TDS in a socket; (b) scan of QR codes; (c) tightening the socket with a torque wrench; (d) test operation via the software GUI.

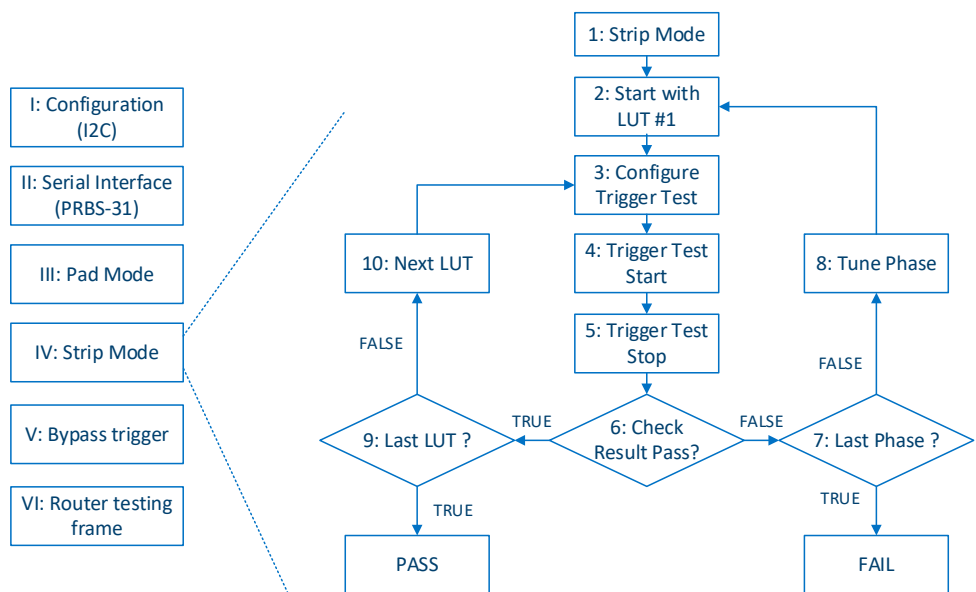

Fig. 7. The flow chart of the test platform software for the Strip Mode test.

I.   Configuration: verification of the I2C configuration interface and read/write of registers;
II.  Serial Interface: evaluation of the performance of the 4.8 Gb/s interface with embedded PRBS-31 serial pattern;
III. Pad Mode: evaluation of the pad mode of the TDS;
IV.  Strip Mode: evaluation of the strip mode of the TDS;
V.   Bypass trigger: a test function bypassing the external trigger interface;
VI.  Router testing frame: fake strip mode data pattern generator.

The six tests cover all functions of the TDS, and verification of each test follows pre-assigned testing procedures. An illustration of the flow chart for Test IV is shown in Fig. 7. There are a total of 10 steps which start from acknowledging a station with the specific testing type and configurations in steps 1-3. Once the station is ready, tests for the strip mode will be executed, results checked in the VC707 FPGA, and status codes sent back to the software, as shown in steps 4-6. A thorough evaluation of the strip mode was performed by a total of sixteen pre-assigned LUTs. Each LUT includes a channel offset and a trigger operation covers a range of 14 strips from the offset channel in its LUT. By selecting different channel offsets for the LUTs, sixteen LUTs are adequate to cover all 128 strip channels in an evaluation. Tests with the LUTs are executed in series and a "PASS" flag is given once all tests are completed, as shown in steps 9-10. The trigger information with each LUT is sent from the VC707 to its mezzanine card via a 640 Mb/s interface as shown in Fig. 2. Phase alignment of the decoding clock with respect to its data lines is important and is tuned for an optimal setup/hold time in case failures are observed in step 6. The phase is tuned in steps of about 78 ps through the FPGA. The test continues when a working phase is found, otherwise the test is marked as "FAIL" when failures persist over a scanning of one serial bit width (640 Mb/s: 1.5625 ns). The phase scanning is introduced to exclude possible failures from timing violations in the trigger interface. Similar flow charts are followed for all other tests.

A full evaluation of all six tests for a TDS takes about 15 minutes. Since the platform is scalable and we have three stations cascaded, the average testing time per chip is reduced to about 5 minutes. Tests are performed by the platform automatically and the only user intervention is listed in Fig. 6. For the mass production, about 6,000 TDS chips were produced and it took about three months to complete the evaluations for all chips. The test pass rate was found to be over 95%.

## 4.   Conclusion

The TDS ASIC is a critical component in the frontend electronics for the NSW upgrade of the ATLAS muon spectrometer. It is a mixed-signal chip handling up to 128 detector channels and has various modes to support the readout of both pads and strips of the sTGC detector. Production verification of TDS requires thorough characterization of each ASIC. We designed an automatic test platform to provide quality control of all production ASICs. The platform has a three layer implementation with custom hardware, firmware, and software. The system is scalable and multiple testing stations were cascaded to speed up the evaluation progress. With three stations cascaded, a total of about 6,000 TDS in production were qualified in three months, with a successful pass rate over 95%. The implementation can be a reference to similar applications that require custom test platforms for qualification of custom ASICs in mass production.

**Acknowledgments**

This works is supported by the Department of Energy under contract DE-AC02-98CH10886. The authors would like to thank Edward Diehl from the University of Michigan for his help in this work.

REFERENCES

[1] ATLAS New Small Wheel Technical Design Report, document CERN-LHCC-2013-006 and ATLAS-TDR-20-2013, Jun. 2013.
[2] A. Abusleme, C. Belanger-Champagne, A. Bellerive, *et al*, "Performance of a full-size small-strip thin gap chamber prototype for the ATLAS new small wheel muon upgrade" Nucl. Inst. And Methods in Physics Research, A. vol. 817, no. 1, pp. 85-92, May 2016.
[3] G. De Geronimo, J. Fried, S. Li, et al. "VMM1- An ASIC for Micropattern Detectors" IEEE Trans. On Nuclear Sci. vol. 60, no.3. pp. 2314-2321, May 2013.
[4] J. Wang, L. Guan, J. W. Chapman, et al "Design of a Trigger Data Serializer ASIC for the Upgrade of the ATLAS Forward Muon Spectrometer", IEEE Trans. On Nuclear Sci. vol. 64, no.12. pp. 2958-2965, Dec. 2017.
[5] J. Wang, L. Guan, J. Chapman et al., "A programmable time alignment scheme for detector signals from the upgraded muon spectrometer at the ATLAS experiment," Nucl. Inst. And Methods in Physics Research, A. vol. 871, no. 1, pp. 8-12, Nov. 2017.
[6] J. Wang, X. Hu, T. Schwarz et al "FPGA Implementation of a Fixed Latency Scheme in a Signal Packet Router for the Upgrade of ATLAS Forward Muon Trigger Electronics", IEEE Trans. On Nuclear Sci. vol. 62, no.5. pp. 2192-2201, Oct. 2015.
[7] J. Wang, X. Hu, R. Pinkham et al "Fixed-Latency Gigabit Serial Links in a Xilinx FPGA for the Upgrade of the Muon Spectrometer at the ATLAS Experiment", IEEE Trans. On Nuclear Sci. vol. 65, no.1. pp. 656-664, Jan. 2018.
[8] J. Wang, L. Guan, Z. Sang et al "Characterization of a Serializer ASIC chip for the upgrade of the ATLAS muon detector", IEEE Trans. On Nuclear Sci. vol. 62, no.6. pp. 3242-3248, Dec. 2015.
[9] https://www.ironwoodelectronics.com/Catalog/Content/Templates/PartGrids.cfm?StartRow=1&cPart=SG-BGA-6052&Grid=SG-BGA_TABLE-1mm.